\documentclass{article}

\usepackage{microtype}
\usepackage{graphicx}
\usepackage{dcolumn}
\usepackage{bm}
\usepackage{times}
\usepackage{outlines}
\usepackage{amsmath}
\usepackage{amsthm}
\usepackage{amsfonts}
\usepackage{ifthen}

\usepackage{hyperref}
\ifdefined\XeTeXrevision\else\usepackage{breakurl}\fi

\usepackage[accepted]{mlsys2025}
\makeatletter\renewcommand{\Notice@String}{}\makeatother

\let\cite\citep

\usepackage{ulem}
\usepackage{xspace}
\usepackage{paralist}
\usepackage[dvipsnames]{xcolor}
\usepackage{enumitem}
\usepackage{outlines}
\usepackage{subcaption}
\usepackage{booktabs}
\usepackage{fnpct}
\usepackage{xparse}
\usepackage{xstring}
\usepackage{cleveref}
\crefname{figure}{Figure}{Figures}
\crefname{table}{Table}{Tables}
\crefformat{section}{\S#2#1#3}
\Crefformat{section}{\S#2#1#3}
\crefformat{subsection}{\S#2#1#3}
\Crefformat{subsection}{\S#2#1#3}
\crefformat{subsubsection}{\S#2#1#3}
\Crefformat{subsubsection}{\S#2#1#3}

\ExplSyntaxOn
\NewDocumentCommand{\btexttt}{m}
 {
  \texttt{
    \tl_set:Nn \l_tmpa_tl {#1}
    \tl_map_function:NN \l_tmpa_tl \__insert_breakpoints:n
  }
 }

\cs_new:Nn \__insert_breakpoints:n
 {
  \str_if_in:nnTF {./-} {#1}
   {#1\hspace{0pt}}
   {#1}
 }
\ExplSyntaxOff

\newcommand{\hide}[1]
{
\ifthenelse{\boolean{false}}{#1}{}
}
\newcommand{\secref}[1]{\S\ref{#1}}
\newtheoremstyle{mystyle} 
  {.3\topsep}               
  {.3\topsep}               
  {\itshape}              
  {}                      
  {\bfseries}             
  {.}                     
  {.5em}                  
  {}                      

\theoremstyle{mystyle}

\newcommand{\eat}[1]{}
\usepackage{multirow}
\usepackage{subcaption}
\usepackage{pifont}
\usepackage{enumitem}
\usepackage{refcount}
\setlist[itemize]{itemsep=0pt, partopsep=0pt, parsep=1pt, topsep=1pt}
\setlist[enumerate]{itemsep=0pt, partopsep=0pt, parsep=1pt, topsep=1pt}

\usepackage{booktabs} 
\usepackage{siunitx} 
\definecolor{darkkhaki}{RGB}{189,183,107}
\definecolor{teal}{RGB}{0,128,128}
\definecolor{orange}{RGB}{230,126,34}

\newenvironment{packeditemize}{\begin{list}{$\bullet$}{\setlength{\itemsep}{0.5pt}\addtolength{\labelwidth}{-4pt}\setlength{\leftmargin}{2ex}\setlength{\listparindent}{\parindent}\setlength{\parsep}{1pt}\setlength{\topsep}{2pt}}}{\end{list}}

\Crefname{section}{Sec.}{Sec.}
\Crefname{algorithm}{Alg.}{Alg.}
\Crefname{figure}{Fig.}{Fig.}

\newcommand{\sys}{\btexttt{MORI}\xspace}

\newcommand{\MyPara}[1]{\vspace{1mm}\textbf{\textit{#1}}~}

\mlsystitlerunning{Idleness Is Relative}

\begin{document}

\twocolumn[
\mlsystitle{Idleness Is Relative: Exploiting Tool-Call Idle Windows for Offloading in Agentic Systems with MORI}

\mlsyssetsymbol{equal}{*}

\begin{mlsysauthorlist}
\mlsysauthor{Tian Xia}{equal,ucb}
\mlsysauthor{Yi Xu}{equal,ucb}
\mlsysauthor{Hanchen Li}{ucb}
\mlsysauthor{Zhifei Li}{pr}
\mlsysauthor{Xiaokun Chen}{stf}
\mlsysauthor{Hao Kang}{git}
\mlsysauthor{Yifan Qiao}{ucb}
\mlsysauthor{Ion Stoica}{ucb}
\end{mlsysauthorlist}

\mlsysaffiliation{ucb}{University of California, Berkeley.}
\mlsysaffiliation{pr}{Princeton University.}
\mlsysaffiliation{stf}{Stanford University.}
\mlsysaffiliation{git}{Georgia Institute of Technology.}

\mlsyscorrespondingauthor{Tian Xia}{tianxia@berkeley.edu}

\mlsyskeywords{LLM Inference, Agents, Memory Offloading, Scheduling}

\vskip 0.3in

\begin{abstract}
Agentic programs alternate between LLM inference and tool calls, each request extending the previous one's context. Program-aware schedulers pin a program's KV cache in GPU HBM across tool calls to avoid re-prefilling it at every step, but as contexts grow, pinned KV cache exceeds HBM, raising \emph{program eviction}: deciding when a program releases GPU memory and when it returns. Offloading to CPU DRAM lets an evicted program wait out a tool call instead of recomputing its context, but existing schedulers leave DRAM to LRU, so offloaded programs thrash. Making these decisions is hard: tool-call durations vary by orders of magnitude and are unknown in advance. We observe that programs nevertheless stay busy (short tool calls) or idle (long ones) for extended periods, and use a program's idleness, its recent fraction of time in tool calls, as a continuous, relative metric. \sys{} ranks programs by idleness, keeps the least idle in HBM and the next in DRAM, and drops the rest, enforcing admission control at both tiers under any GPU-to-CPU capacity ratio; it also proactively demotes idle programs to make room for ready ones. On real coding-agent traces, \sys{} achieves on average 64\% higher throughput and 72\% lower p90 TTFT than the best offloading baseline under high load.
\end{abstract}
]

\printAffiliationsAndNotice{\mlsysEqualContribution}  

\section{Introduction}

\begin{figure*}[t]
  \centering
  \includegraphics[width=\textwidth]{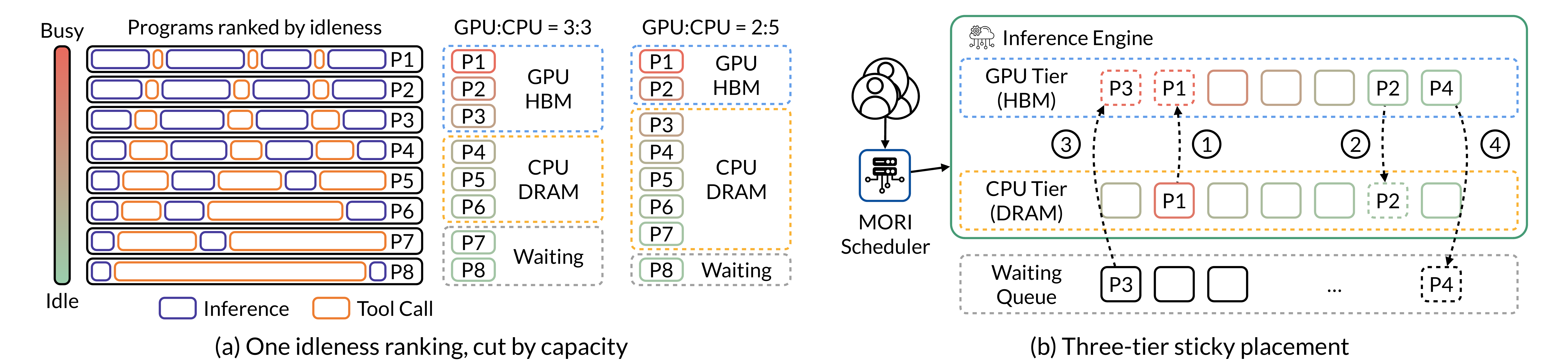}
  \caption{Overview of \sys{}. (a)~\sys{} ranks programs by idleness and splits the ranking by tier capacity, counted in programs: the least idle programs stay in GPU HBM, the next are offloaded to CPU DRAM, and the rest wait in neither tier, so a different GPU:CPU capacity only moves the split. (b)~\sys{} moves a program when a tier overflows, GPU capacity frees up, or a ready program does not fit. Programs are colored by idleness as in (a) (black when Waiting), and dashed boxes mark where a program moves: \ding{172}~Reload, \ding{173}~Offload, \ding{174}~New, and \ding{175}~Pause; \ding{172}~with~\ding{173} is a proactive promotion.}
  \label{fig:overview}
\end{figure*}

Modern large language model (LLM) serving systems increasingly execute long-running agentic workloads~\cite{anthropic2025claudecode, openai2025codex, anysphere2024cursor, wang2025openhands} rather than isolated chat requests.
In these workloads, a program is not a single prompt-response interaction but a stateful execution that follows a ReAct-style loop~\cite{yao2023react, shinn2023reflexion} of model reasoning and tool-call actions: each model invocation produces an action, the action's result is appended to the context, and the next invocation continues from the extended state.
As a result, the unit of serving is no longer an individual LLM request, but an agent program whose context, tool state, and memory footprint accumulate over time.

Conventional serving systems~\cite{kwon2023vllm, zheng2024sglang}, however, treat the requests of the same program as disjoint: once a request completes, its KV cache becomes a candidate for block-level eviction policies such as LRU, even though the program's next request will reuse it. As a result, during a short gap induced by a tool call, a program might lose its KV cache to other programs' requests and re-prefill it at every step. Recent program-aware schedulers~\cite{kang2026thunderagent, li2025continuum} instead recognize requests as steps of one program and pin its KV cache in GPU HBM across tool calls, avoiding this thrashing.

As programs' contexts grow, however, the aggregate KV cache of concurrent programs quickly exceeds GPU capacity, so a program cannot stay pinned until it finishes: a program blocked on a long tool call holds a large KV cache while idle, wasting HBM that other programs could use. Pinning therefore introduces a new type of eviction, \emph{program eviction}: deciding \emph{when} a program stops reserving GPU memory, as opposed to the engine's block-level \emph{KV cache eviction} (e.g., LRU), which only picks \emph{which} blocks to drop once memory is full. The reverse decision, \emph{program promotion}, matters just as much: when an evicted program's tool call completes, the scheduler decides when to bring it back to the GPU, and whether to proactively evict another program to make room for it.

KV cache offloading~\cite{sheng2023flexgen, gao2024cachedattention, liu2025lmcache, qin2024mooncake} gives an evicted program a second tier: its KV cache can wait in CPU DRAM during a tool call instead of being recomputed. DRAM is finite too, however. Existing program-aware schedulers manage only HBM: combined with a KV cache offloading layer, they inherit its default LRU policy, so every evicted program is offloaded, and offloaded programs evict one another and thrash in DRAM (\cref{sec:eval-why}), causing recomputation when they resume. DRAM thus needs admission control as well, so for each evicted program the scheduler must also decide whether to keep its KV cache in DRAM or to drop it and recompute it later.

These decisions are hard to make well. Chat serving systems can offload a conversation's KV cache after every turn, because a human takes tens of seconds to reply (\cref{sec:mot-variance}), but many agentic tool calls finish within hundreds of milliseconds, often faster than a program's long context can be moved out and back. A program should therefore stay in HBM through short tool calls and leave only for long ones, yet tool-call durations vary by orders of magnitude, even for the same tool, and are unknown until the call completes. Existing program-aware schedulers have limitations under this unpredictability (\cref{sec:mot-variance}). Continuum~\cite{li2025continuum} predicts each call's duration from past calls of the same tool and keeps the program pinned for a time-to-live (TTL) derived from that prediction, so a misprediction either releases a program shortly before it resumes or keeps an idle one pinned. ThunderAgent~\cite{kang2026thunderagent} evicts a program only once GPU capacity is full and selects the victim by context length, so an idle program keeps its HBM while programs whose tool calls have completed queue behind it, and the victim may be about to resume (\cref{sec:eval-main}).

Our key observation from trace analysis is that each agentic program typically stays in one of two phases for an extended period: a \textit{busy} phase of short tool calls on the order of hundreds of milliseconds (e.g., reading or writing files), and an \textit{idle} phase blocked on a long tool call lasting tens of seconds or longer (e.g., waiting for human input or a spawned subagent; \cref{sec:mot-phases}). Naturally, placing busy-phase programs on HBM and offloading idle-phase programs to DRAM aligns each tier with its expected access pattern. However, a binary busy/idle classification is insufficient: the amount of KV cache held by busy versus idle programs shifts over time and need not match the GPU-to-CPU capacity ratio, so when busy programs exceed HBM, or idle programs exceed DRAM, the scheduler must still decide which ones to keep.

Instead, we use a program's \textbf{idleness}, the fraction of its recent time spent in tool calls, as a continuous, relative metric: rather than labeling programs busy or idle, a scheduler can rank them by how idle they are and let each tier's capacity decide where to cut the ranking. We propose \sys{} (\Cref{fig:overview}; \textbf{M}emory \textbf{O}ffloader with \textbf{R}elative \textbf{I}dleness), which computes idleness over a window of recent steps (\cref{sec:des-metric}) that reacts quickly to phase transitions while smoothing outliers, so \sys{} anticipates a program's behavior from its own recent history rather than from a model of individual tool-call durations. Following this ranking, HBM holds the least idle programs, DRAM the next, and the rest are dropped to avoid thrashing, which enforces admission control at both tiers and adapts to any GPU-to-CPU capacity ratio without per-hardware tuning. When a tier overflows, \sys{} \emph{demotes} the most idle programs; when GPU capacity frees up, it \emph{promotes} the least idle \emph{ready} programs, those whose tool call has completed. Unlike prior schedulers, \sys{} also promotes \emph{proactively}: when a ready program does not fit, \sys{} uses a current-cycle form of idleness to decide whether to move some GPU-resident programs to DRAM to make room for it. Otherwise, placement is \emph{sticky} to avoid repeated offloads and reloads (\cref{sec:des-policy}).

We implement \sys{} on top of ThunderAgent~\cite{kang2026thunderagent} and SGLang~\cite{zheng2024sglang} and evaluate it with three models (Qwen2.5-7B and gpt-oss-120b on H200, Qwen3-235B on B300) on three coding-agent workloads built from real developers' sessions: Claude Code (CC) sessions with subagents from AgentX~\cite{semianalysis2026agentx}, and CC and Codex sessions from TraceLab~\cite{tracelab2026}. Under high load, \sys{} achieves on average 64\% higher throughput and 72\% lower p90 TTFT than the best baseline across the three workloads. \sys{} also adapts well to different GPU-to-CPU capacity ratios: with the same total KV capacity, \sys{} delivers 24\% more throughput than the best baseline while moving 60\% of the GPU KV memory to CPU (GPU:CPU = 1:4 vs.\ 1:1; \cref{sec:eval-ratio}).

In summary, we make the following contributions:
\begin{packeditemize}
  \item We characterize real agentic workloads on public traces from TraceLab and AgentX, and show that despite highly variable tool-call durations, programs stay in busy or idle phases for extended periods (\cref{sec:motivation}).
  \item We design \sys{}, which ranks programs by a continuous, windowed idleness metric to place them across HBM and DRAM with admission control at both tiers, promotes ready programs proactively, and adapts to different GPU-to-CPU memory ratios without per-hardware tuning (\cref{sec:methodology}).
  \item We evaluate \sys{} against production and research systems on three model and GPU configurations and three coding-agent workloads, demonstrating higher throughput and lower tail latency (\cref{sec:evaluation}).
  \item We open-source \sys{} to stimulate further research in serving agentic workloads.\footnote{\url{https://github.com/cblmemo/mori}}
\end{packeditemize}

\section{Background}
\label{sec:background}

\subsection{KV Cache Eviction and Program Eviction}
\label{sec:bg-kvcache}

LLM inference engines such as vLLM~\cite{kwon2023vllm} and SGLang~\cite{zheng2024sglang} cache the key and value projections of processed tokens as the \emph{KV cache}, which grows linearly with context length and dominates GPU memory for long contexts. After a request completes, its KV cache is kept as a prefix cache so that later requests sharing the prefix skip recomputation; when memory runs out, cached blocks are evicted, typically least recently used first. This \emph{KV cache eviction} decides \emph{which} blocks to drop.

\begin{figure}[t]
  \centering
  \includegraphics[width=0.9\linewidth]{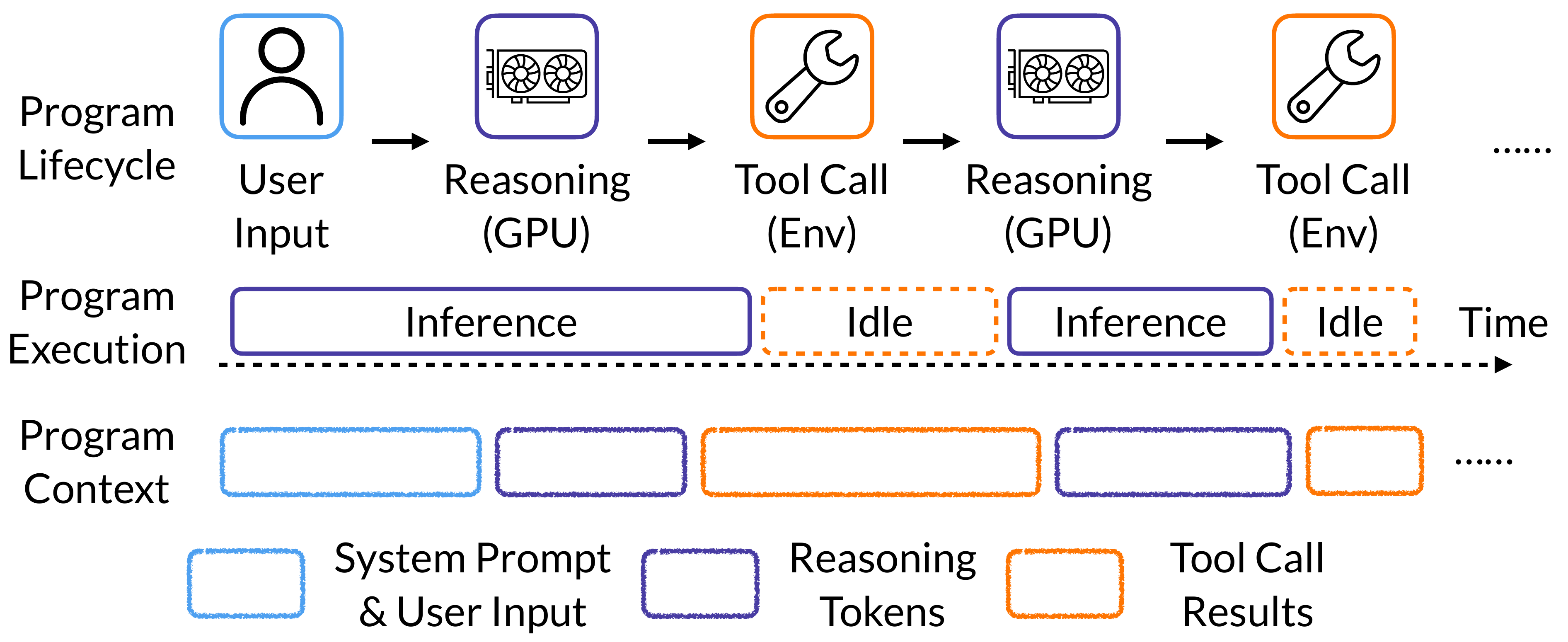}
  \caption{Structure of an agentic program: its lifecycle (top) alternates inference on the GPU with tool calls in the environment, its execution (middle) is idle during each tool call, and its context (bottom) grows with every step.}
  \label{fig:program}
\end{figure}

In agentic workloads, a single user task produces a chain of model invocations connected by prefix dependencies. We define an \emph{agentic program} (or simply \emph{program}) as the complete sequence of model invocations that constitute a single agent session, together with the tool calls that interleave them (\Cref{fig:program}). A program consists of alternating \emph{steps}, each comprising one inference call on the GPU followed by a \emph{tool-call gap} where the program is waiting for an external action to complete; parallel tool calls form a single gap whose duration is the maximum of the individual durations. Each request extends the context of its previous request, so the KV cache grows monotonically across steps and remains valuable during tool-call gaps, when the model is idle. A subagent~\cite{wu2023autogen} is an independent program with its own context; from its parent's perspective, the subagent's execution is a single, potentially long tool call.

These request-level systems cannot tell a program in a short tool-call gap from one that has finished, so concurrent programs evict each other's KV caches and re-prefill them at every step (\cref{sec:eval-why}). LRU can even invert the right decision: a program in a short tool call that started a few seconds ago looks less recently used than one that has just entered a tool call lasting tens of seconds, so LRU evicts the former, which will resume shortly, and retains the latter.

Program-aware scheduling identifies the requests of a program and pins its KV cache in GPU HBM across tool calls, so that the program's next request reuses it instead of re-prefilling the context. Pinning introduces a second, scheduler-level decision, \emph{program eviction}: deciding \emph{when} a program stops reserving GPU memory so that its KV cache becomes evictable again. Releasing a program too early brings back the thrashing that pinning prevents, while releasing it too late lets an idle program hold HBM that other programs could use.

A demoted program is \emph{ready} once its tool call has completed and its next request waits at the scheduler. The reverse decision is \emph{program promotion}: deciding \emph{when} a ready or new program regains GPU memory.

\subsection{KV Cache Offloading}
\label{sec:bg-offloading}

Without a second tier, an evicted program loses its KV cache and recomputes it when it resumes. Host DRAM offers a larger second tier: a typical H100 server pairs eight 80\,GB GPUs with 1\,TB of DRAM~\cite{nvidia2023h100whitepaper}, about 1.6$\times$ the aggregate HBM. KV cache offloading moves KV cache that is not immediately needed to DRAM and loads it back on demand, replacing recomputation with a transfer~\cite{xie2025hicache, liu2025lmcache}. Unlike the short requests for which recomputation was found competitive with swapping~\cite{kwon2023vllm}, agentic contexts reach tens of thousands of tokens and are reused at every step; since prefill cost grows quadratically with context length while transfer cost grows linearly, reloading a long prefix from DRAM is several times cheaper than recomputing it~\cite{pan2025kvflow, gao2024cachedattention, qin2024mooncake}.

\section{Motivation}
\label{sec:motivation}

\subsection{Offloading and DRAM Thrashing}
\label{sec:mot-dram}

KV cache offloading lets an evicted program wait in CPU DRAM instead of recomputing its KV cache (\cref{sec:bg-offloading}), but DRAM is finite too. Existing program-aware schedulers manage only GPU HBM; combined with a KV cache offloading layer, they inherit its default LRU policy and offload every evicted program. Once DRAM fills, offloaded programs contend for it and evict one another (\Cref{fig:eval-cache}(b)), causing reload thrashing analogous to GPU-side thrashing~\cite{kang2026thunderagent, li2025continuum}, so admission control is needed at both tiers. Bounding DRAM is thus program eviction at the CPU tier: rather than letting the engine's LRU drop parts of many offloaded programs' caches, the scheduler decides which programs keep their whole cache in DRAM. A program that does not fit is not admitted to DRAM, instead of displacing those already there, and recomputes its KV cache when it resumes. Both decisions, when a program leaves HBM and whether its KV cache goes to DRAM, are complicated by the high variance in tool-call durations.

\subsection{Tool-Call Variance and Existing Program Eviction}
\label{sec:mot-variance}

\begin{figure}[t]
  \centering
  \includegraphics[width=\linewidth]{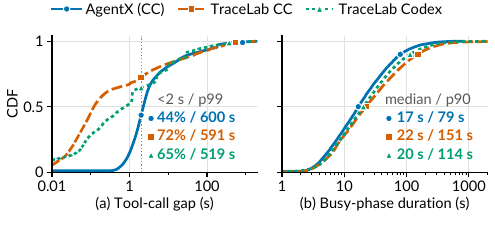}
  \caption{(a)~CDF of tool-call gap durations and (b)~CDF of wall-clock busy-phase durations (2\,s threshold) in AgentX's Claude Code (CC) traces~\cite{semianalysis2026agentx} and in TraceLab's CC and Codex traces~\cite{tracelab2026}. Gaps span several orders of magnitude, from milliseconds (file I/O) to minutes (human input, subagents), yet short tool calls cluster into busy phases lasting tens of seconds.}
  \label{fig:traces}
\end{figure}

We analyze 11.4k sessions from TraceLab~\cite{tracelab2026}, a public corpus of CC and Codex sessions (6.7k and 4.7k, respectively), and 391 CC sessions from AgentX~\cite{semianalysis2026agentx}, whose main agents and subagents form 1.8k programs (\cref{sec:eval-setup}). \Cref{fig:traces}(a) shows the distribution of their tool-call gap durations, which spans several orders of magnitude, from milliseconds for file reads to minutes for human input or subagent invocations. This variance exists not only across tool types but within the same type: a shell command may complete in hundreds of milliseconds or a \texttt{pytest} may run for minutes.

This variance makes offloading decisions difficult. Offloading systems designed for chatbots~\cite{gao2024cachedattention} can offload after every turn because human inter-turn gaps last tens of seconds. Agentic programs break this assumption in two ways: their tool-call gaps are much shorter, and their contexts, which accumulate tool outputs over tens of steps, are much longer (medians of 98k--124k tokens in TraceLab and AgentX), so each transfer moves more data in less time. Moving a program's KV cache between GPU HBM and CPU DRAM costs a transfer linear in its size: for a short tool call the transfer can exceed the gap itself, whereas for a long one it is negligible and keeping the cache in HBM wastes memory that could serve other programs. The right decision depends on each gap's duration, which is unknown until the call completes and depends on external factors invisible to the serving system~\cite{abhyankar2024inferceptefficientinterceptsupport}.

Existing program-aware schedulers have limitations in deciding program eviction under this variance. Continuum~\cite{li2025continuum} predicts each call's duration from past calls of the same tool, pins the program only when the call is expected to be short, and unpins it when a time-to-live (TTL) expires; a misprediction either releases a program shortly before it resumes or keeps an idle one pinned, and calls with heavy-tailed durations are rarely pinned at all (\cref{sec:eval-main}). ThunderAgent~\cite{kang2026thunderagent} avoids prediction but evicts passively: a program loses its residency only when GPU capacity overflows, and the victim is chosen by context length rather than tool-call behavior, so an idle program holds HBM while ready programs queue, and the victim may be about to resume (\cref{sec:eval-main}). Instead of predicting individual calls, we look for exploitable patterns at a coarser, program-level granularity.

\subsection{Busy and Idle Phases}
\label{sec:mot-phases}

While individual tool-call durations vary widely, agentic programs exhibit a structured \emph{phase behavior} over time. Through trace analysis, we identify two alternating phases:

\MyPara{Busy phase.} The program interacts with its environment through a sustained sequence of short tool calls (reading files, writing edits, issuing lightweight shell commands), each completing in hundreds of milliseconds to a few seconds, and returns to the inference engine after each gap.

\MyPara{Idle phase.} The program is blocked on a long-running tool call (executing a test suite, waiting for human approval, or awaiting a subagent's result) for tens of seconds to minutes, and keeping its KV cache in GPU memory brings no immediate benefit.

We quantify this phase structure in \Cref{fig:traces}(b). We define a busy phase as a maximal run of steps separated by tool calls no longer than a 2\,s threshold, and measure its wall-clock duration, which includes both the inference steps and the short tool calls in the run. Busy phases last a median of 17\,s, 22\,s, and 20\,s in AgentX, TraceLab CC, and TraceLab Codex, respectively (p90 of 79\,s, 151\,s, and 114\,s). At the 2\,s threshold, 44\% (AgentX), 72\% (TraceLab CC), and 65\% (TraceLab Codex) of gaps are short, while long gaps, which include human waits and, in AgentX, waits for subagents, account for 99\% of gap time. Short tool calls thus cluster into sustained bursts lasting tens of seconds, while a small fraction of long gaps dominates idle time. This structure generalizes across models (multiple versions of Claude and GPT), agent harnesses (CC and Codex), and a variety of real-world workloads~\cite{tracelab2026}.

\begin{figure}[t]
  \centering
  \includegraphics[width=\linewidth]{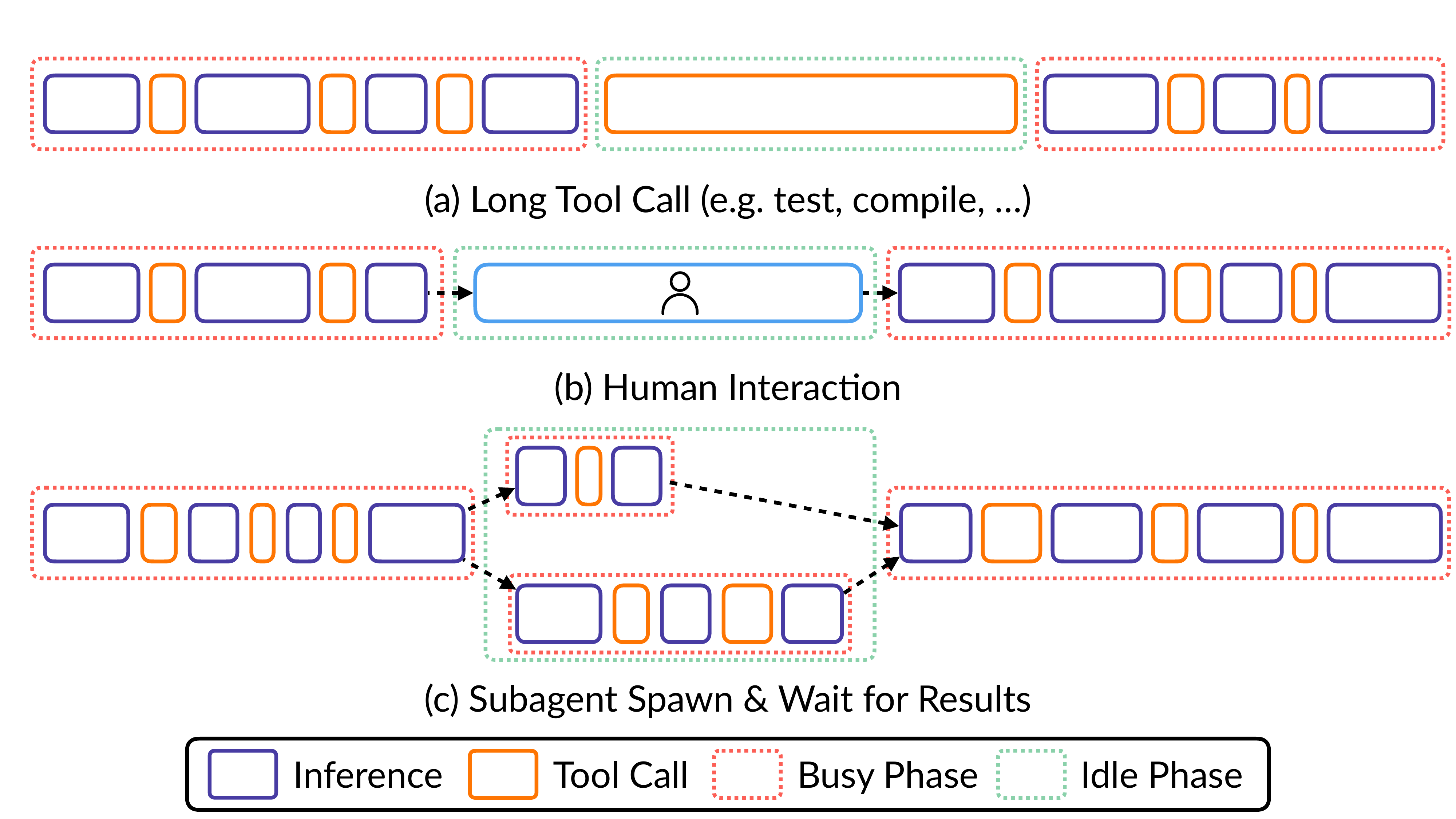}
  \caption{Three examples of idle phases in agentic programs: (a)~a long-running tool call such as a test suite or compilation, (b)~waiting for human interaction (e.g., approval or feedback), and (c)~spawning subagents and waiting for their results. Red dashed frames denote busy phases; green dashed frames denote idle phases. Note that in~(c), the subagents themselves may be in busy phases, but the parent agent remains idle until they return.}
  \label{fig:phases}
  \vspace{-16pt}
\end{figure}

Two properties make phases a robust signal for scheduling. First, phase transitions are \emph{infrequent}. The same span of tens of seconds can hold many short calls, forming a busy phase (\cref{fig:traces}(b)), or a single long call, forming an idle phase (\cref{fig:phases}); such calls reach a p99 of 519--600\,s (\cref{fig:traces}(a)). Once a program enters a phase, it typically stays there for an extended period of time (tens of seconds). Second, a program's current phase is \emph{identifiable} from recent history without predicting future tool-call durations: a program that has issued several short tool calls in succession is likely in a busy phase, while a program whose current tool call has already run significantly longer than its recent calls is likely entering an idle phase. In our traces, a gap that follows a short gap is itself short 75--81\% of the time, and a call that has already lasted as long as the inference step before it is 2.6--4.5$\times$ as likely to run for another 10\,s as an average call is to last 10\,s. These properties let the scheduler reliably place each program in a memory tier based on its recent behavior, which \sys{}'s idleness metric captures (\cref{sec:des-metric}).

\subsection{Mapping Phases to the Memory Hierarchy}
\label{sec:mot-hierarchy}

The two-phase structure maps naturally to the two-tier memory hierarchy: busy-phase programs will resume inference shortly and should keep their KV caches in HBM, while idle-phase programs will not need theirs in the near term and can be offloaded to DRAM. A fixed busy/idle partition, however, rarely matches the available capacity. Programs continuously switch phases and their contexts grow, so the aggregate memory demand of busy and idle programs changes over time, while a server's GPU-to-CPU capacity ratio is fixed and differs across configurations (e.g., 1:1.6 or 1:3.1 for an 8$\times$H100 node with 1\,TB or 2\,TB of DRAM).

If more programs are busy than HBM can hold, some busy programs must be evicted: the \emph{most busy} ones should keep HBM, since their KV caches sit idle there for the least time, and the less busy ones should be offloaded. If fewer are busy, HBM is wasted (\cref{sec:eval-ratio}) unless the idle programs that will resume inference sooner are prioritized for it, while the \emph{most idle} ones, i.e., most likely to remain in their tool calls for the longest time, should be placed in CPU DRAM. With DRAM also bounded (\cref{sec:mot-dram}), the same question arises at the CPU tier: which idle programs keep DRAM. The key question is thus not whether a program is idle or busy in absolute terms, but whether it is \emph{more idle} than others. \Cref{sec:methodology} presents how \sys{} leverages this idea.

\section{Design}
\label{sec:methodology}

\subsection{Overview}
\label{sec:des-overview}

\sys is a program-aware scheduler that sits between agent clients and an LLM inference engine (\Cref{fig:overview}(b)). Each agent client annotates its requests with a \emph{program ID}~\cite{li2025continuum, kang2026thunderagent, luo2025autellix}, which \sys uses to track each program across its entire lifetime of inference steps and tool calls. \sys proxies every request, places programs across memory tiers (\cref{sec:des-policy}), and provides hints to the engine for KV cache placement (\cref{sec:des-typed}).

\MyPara{Three tiers.} \sys organizes programs into three tiers, each corresponding to a level of the memory hierarchy:

\begin{itemize}
  \item \textbf{GPU tier (HBM).} Its programs' KV caches reside in GPU HBM, and their requests are forwarded directly to the inference engine.
  \item \textbf{CPU tier (DRAM).} Its programs' KV caches are offloaded to CPU DRAM on the same node and must be reloaded to GPU HBM before inference can proceed.
  \item \textbf{Waiting queue.} Its programs reserve memory in neither tier, so their KV caches are evicted first, and promotion back to the GPU typically requires recomputing the KV cache via a full prefill.
\end{itemize}

Requests of programs in the CPU tier or the Waiting queue are gated until the scheduler promotes the program to the GPU tier. The scheduler admits a program to a tier only if its KV cache fits within the tier's remaining capacity; rather than predicting output lengths, it reserves a fixed buffer per GPU-tier program for the tokens a step generates. Bounding both tiers is the admission control that \cref{sec:mot-dram} calls for.

\MyPara{One ranking, cut by capacity.} Following \cref{sec:mot-hierarchy}, \sys ranks programs by idleness (\cref{sec:des-metric}) and cuts the ranking by tier capacity (\Cref{fig:overview}(a)): the least idle programs fill GPU HBM, the next fill CPU DRAM, and the KV caches of the rest are discarded to make room for the other two tiers and recomputed when those programs resume. A different GPU-to-CPU capacity ratio only moves the cut. Re-sorting the programs on every tick, however, would move a program whenever its rank changes, and tier transitions are not free: moving a program from GPU to CPU incurs an offload, and moving it back incurs a reload. Placement is therefore \emph{sticky} (\cref{sec:mot-phases}): a program stays in its tier until the scheduler must move one, when a tier overflows, when GPU capacity frees up, or when a ready program does not fit, and only then does the scheduler consult the ranking (\cref{sec:des-policy}).

\subsection{Idleness Metric}
\label{sec:des-metric}

A program alternates between two statuses, the two parts of a step (\cref{sec:bg-kvcache}): \emph{Reasoning}, when it is actively executing on a GPU, and \emph{Acting}, when it is waiting on an external tool call and its KV cache sits idle. Each incoming request transitions a program to Reasoning; each completed response transitions it back to Acting. A program's status is instantaneous and changes on every request boundary, whereas its phase (\secref{sec:mot-phases}) is a sustained pattern: a busy-phase program rapidly alternates between both statuses, while an idle-phase program remains Acting for an extended period.

\MyPara{Windowed idleness.} For each program, \sys tracks the time spent in Reasoning (GPU inference) and Acting (tool-call gaps) over a sliding window of the most recent $k$ steps. Conceptually, this captures the recent behavior of a program, and since phase transitions are infrequent (\secref{sec:mot-phases}), recent behavior serves as a reliable proxy for the program's future behavior. Formally, we define the \emph{idleness} of a program as:
\begin{equation}
  \iota = \frac{T_\text{acting}^{(k)}}{T_\text{reasoning}^{(k)} + T_\text{acting}^{(k)}}
  \label{eq:idleness}
\end{equation}
where $T_\text{reasoning}^{(k)}$ and $T_\text{acting}^{(k)}$ are the total time spent in each status over the last $k$ inference--tool-call cycles. Any time a program spends waiting on the scheduler (e.g., held in the CPU tier or the Waiting queue) is excluded from both $T_\text{reasoning}$ and $T_\text{acting}$, so the metric reflects only the program's intrinsic behavior. An idleness close to 1 means a program in the idle phase (most time in tool calls), while an idleness close to 0 indicates a program in the busy phase (most time spent in inference).

The window is \emph{responsive}: when a busy program enters a long tool call, the call's elapsed time soon dominates the window and the idleness rises quickly, and when an idle program resumes, its burst of short calls pushes the long call out of the window, and its consecutive GPU inference further lowers the idleness. It is also less sensitive to outliers: a single unexpectedly long call in a busy phase is diluted by the recent short calls, whereas sustained long calls raise the idleness over successive cycles. A larger window $k$ reacts more slowly but tolerates more outliers; we use $k{=}5$.

\MyPara{Current-cycle idleness.} Proactive promotion (\cref{sec:des-policy}) instead needs the latest information, so it uses \Cref{eq:idleness} over only the current cycle:
\begin{equation}
  \iota^{(1)} = \frac{T_\text{acting}^{(1)}}{T_\text{reasoning}^{(1)} + T_\text{acting}^{(1)}},
  \label{eq:cycle-idleness}
\end{equation}
where $T_\text{reasoning}^{(1)}$ is the duration of the program's last inference step and $T_\text{acting}^{(1)}$ is the elapsed time of its ongoing tool call. Unlike the windowed idleness $\iota$, which smooths over a single long call, $\iota^{(1)}$ reacts to the call in progress.

\subsection{Scheduling Policy}
\label{sec:des-policy}

The scheduler runs a periodic control loop (every 5\,s by default; \cref{sec:implementation}). Each tick first performs proactive promotion, then promotes ready programs, and finally demotes programs from any tier that overflows. Each move corresponds to an arrow in \Cref{fig:overview}(b), and whenever the scheduler must choose among programs, it ranks them by idleness.

\MyPara{Demotion (\ding{173}, \ding{175}).} The admission check (\cref{sec:des-overview}) keeps the GPU tier within capacity by design, but the aggregate KV cache of GPU-tier programs, checked on every tick against current context lengths, can still exceed GPU memory capacity when steps generate more tokens than the reserved buffers hold, or when a request appends a long tool output to its program's context. The scheduler then demotes programs until usage fits, gating demoted programs from the inference engine while preserving the KV caches of the remaining programs. The scheduler prioritizes demoting Acting programs (whose KV caches are idle on GPU) over Reasoning programs (which are actively running inference). Among programs of the same status, it selects the one with the highest idleness $\iota$ first, as it is most likely to have entered or be entering an idle phase. Demoted programs move to the CPU tier only if CPU capacity permits (\ding{173}); otherwise they go directly to the Waiting queue (\ding{175}) instead of evicting other programs' KV caches from DRAM, which avoids DRAM thrashing (\cref{sec:mot-dram}). The CPU tier is bounded the same way; if it overflows, its programs that are not yet ready move to the Waiting queue, most idle first.

\MyPara{Promotion to GPU (\ding{172}, \ding{174}).} When GPU capacity becomes available, the scheduler promotes programs in a priority order: (1)~ready CPU-tier programs (\ding{172}); (2)~ready Waiting-queue programs that have run before; (3)~newly arrived programs, smallest context first (\ding{174}). Within each priority level, the scheduler selects the program with the lowest idleness $\iota$ first, as it is most likely to be in or entering a busy phase and will make the most use of GPU residency.

\MyPara{Proactive promotion (\ding{172} with \ding{173}).} Overflow-driven demotion alone is passive (\cref{sec:mot-variance}): an idle program keeps its HBM until capacity runs out, while ready programs wait behind it. When a ready program does not fit on the GPU, \sys{} therefore decides whether a GPU-resident program is worth evicting for it. A resident program gives way only if it is Acting and its ongoing tool call has already lasted at least as long as the inference step before it (\cref{sec:mot-phases}), i.e., if $\iota^{(1)} \geq \frac{1}{2}$ (\Cref{eq:cycle-idleness}). Comparing each program with its own last step needs no prediction of the call's duration and no workload-specific threshold. Candidates are demoted in decreasing order of $\iota^{(1)}$ until the ready program fits, and at most a fixed fraction of them per tick, which bounds churn.

\subsection{Typed Offloading on the Inference Engine}
\label{sec:des-typed}

The scheduler decides \emph{which} programs belong in each tier, but the inference engine manages the actual KV cache blocks and evicts them on its own when memory runs short. \sys{} therefore has the engine realize the scheduler's placement through its block-level eviction: KV blocks of GPU-tier programs are retained in GPU HBM, and those of CPU-tier programs in CPU DRAM.

\sys{} achieves this by propagating a \emph{type label} from the scheduler to each program's KV cache blocks in the engine. When the scheduler places a program in the GPU tier, it marks the program as \textbf{busy}; when it offloads a program to the CPU tier, it marks it as \textbf{idle}; programs in the Waiting queue are marked \textbf{inactive}. Each request carries its program identifier, and when the engine inserts KV blocks into its prefix cache, it stamps them with the program's type.

The engine's eviction policy remains LRU at its core, but it ranks blocks by type label first and applies LRU only among blocks of the same type. Crucially, the type priority order is \emph{reversed} between the two tiers so that each tier preferentially retains the programs assigned to it:

\begin{itemize}
  \item \textbf{GPU HBM:} evicts inactive first, then idle, then busy, so busy programs' blocks are evicted last.
  \item \textbf{CPU DRAM:} evicts inactive first, then busy, then idle, so idle programs' blocks are evicted last.
\end{itemize}

For example, when the scheduler demotes a program to the CPU tier, it relabels the program idle: its blocks stay in HBM until the engine needs space, are then evicted before any busy block and written back to DRAM, and stay there because DRAM evicts idle blocks last. Prefix blocks shared across programs, typically system prompts, are evicted last in both tiers, so they effectively stay pinned in the engine.

The two tiers are kept exclusive: when a promoted program's blocks are loaded back into HBM, the engine frees their DRAM copies, so DRAM is spent only on programs off the GPU and matches the scheduler's decision.

\section{Implementation}
\label{sec:implementation}

We implemented \sys on top of ThunderAgent~\cite{kang2026thunderagent} and SGLang~\cite{zheng2024sglang} (v0.5.10), adding approximately 3,000 lines of Python to ThunderAgent's scheduler and 1,000 lines to SGLang, mostly in its cache subsystem (via HiCache~\cite{xie2025hicache}).

\MyPara{Scheduler.} \sys exposes a standard OpenAI-compatible chat/completion API augmented with a \texttt{program\_id} field: the agent client tags every request of a program with the same \texttt{program\_id}, and a subagent uses a distinct one. The scheduler runs as an asynchronous control loop (default tick: 5\,s) and retrieves context lengths and memory capacities from the inference engine.

\MyPara{Engine.} Each tick's type-label updates (\secref{sec:des-typed}) are batched into a single request, and SGLang stores the labels on the nodes of its radix-tree prefix cache. HiCache's GPU and host eviction policies order nodes by these labels before LRU, and loading a node back into HBM frees its DRAM copy, which keeps the two tiers exclusive.



\section{Evaluation}
\label{sec:evaluation}

\begin{figure*}[t]
  \centering
  \includegraphics[width=\textwidth]{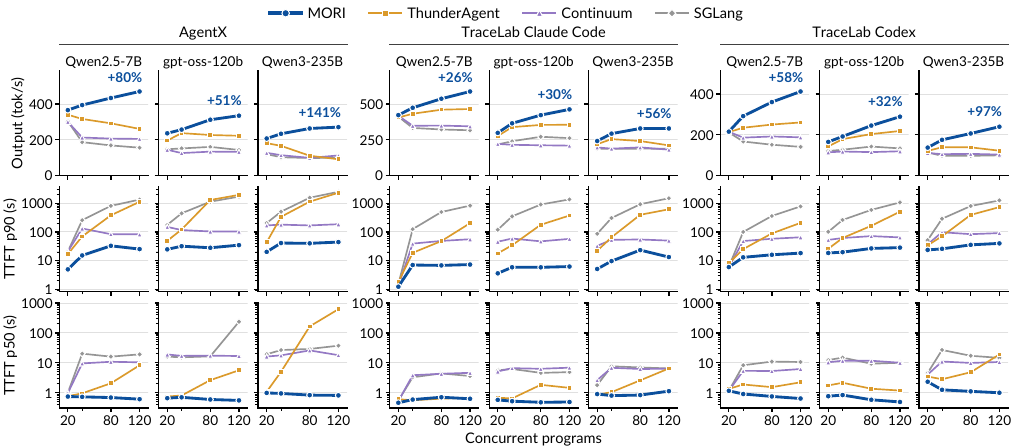}
  \caption{Output throughput (top), p90 TTFT (middle), and median TTFT (bottom), with TTFT on a log scale, at GPU:CPU = 1:1 (configurations in \Cref{tab:eval-configs}). Labels give \sys{}'s throughput gain over the best baseline at the highest concurrency.}
  \label{fig:eval-main}
\end{figure*}

We evaluate \sys{} on three GPU and model configurations and three coding-agent workloads. We ask how much \sys{} improves throughput and latency over existing schedulers with offloading (\cref{sec:eval-main}), where the gains come from (\cref{sec:eval-why}), and how the comparison changes with the GPU-to-CPU capacity ratio (\cref{sec:eval-ratio}).

\begin{table}[h]
  \centering
  \small
  \setlength{\tabcolsep}{4pt}
  \begin{tabular}{llcc}
    \toprule
    GPU & Model & TP & KV pool \\
    \midrule
    H200 & Qwen2.5-7B-Instruct-1M & 1 & 1.94M \\
    H200 & gpt-oss-120b & 1 & 0.71M \\
    B300 & Qwen3-235B-A22B (FP8) & 2 & 1.29M \\
    \bottomrule
  \end{tabular}
  \caption{GPU and model configurations. An H200 has 141\,GB of HBM and a B300 275\,GB. The KV pool is the GPU KV cache capacity in tokens.}
  \label{tab:eval-configs}
\end{table}

\subsection{Setup}
\label{sec:eval-setup}

\MyPara{Hardware and models.} \Cref{tab:eval-configs} lists the configurations, which serve Qwen2.5-7B-Instruct-1M~\cite{qwen2025qwen251m}, gpt-oss-120b~\cite{openai2025gptoss}, and Qwen3-235B-A22B-Instruct-2507~\cite{qwen2025qwen3}; Qwen3-235B uses tensor parallelism (TP) across two B300 GPUs. All systems run on the same SGLang~\cite{zheng2024sglang} build (v0.5.10) and offload to CPU DRAM through HiCache~\cite{xie2025hicache}. Unless noted, the CPU KV pool is as large as the GPU's (GPU:CPU = 1:1); \cref{sec:eval-ratio} varies the ratio.

\MyPara{Workloads.} We replay three coding-agent workloads of real developers' sessions, reproducing the recorded request lengths, tool-call durations, and waits. \emph{AgentX}~\cite{semianalysis2026agentx} consists of 391 Claude Code (CC)~\cite{anthropic2025claudecode} sessions, which we split into 790 programs at idle periods longer than 30 minutes and at context compactions. A quarter of these programs spawn subagents (965 in total), each of which we replay as a program of its own, started and awaited as recorded. Of the main agents' steps, 11\% end in a gap of at least 60\,s (median 3\,min), which covers user input, permission prompts, and slow tools.
We also replay \emph{TraceLab} (CC and Codex)~\cite{tracelab2026}, where about 8\% of steps end in a human wait (median one to two minutes). The median contexts are 108k, 98k, and 124k tokens in AgentX, TraceLab CC, and TraceLab Codex, respectively.
Every replayed request generates exactly its recorded number of output tokens, which include the reasoning tokens of the traced models (Claude Opus~4.6--4.8, Fable~5, and Sonnet~4.6 with thinking, and GPT-5-family models), regardless of the served model.

\MyPara{Baselines.} We compare \sys{} with three baselines, all offloading through HiCache:
\begin{packeditemize}
  \item \textbf{SGLang}: SGLang's native request-level scheduler, which queues requests in the engine and serves those with the longest cached prefix first.
  \item \textbf{ThunderAgent}~\cite{kang2026thunderagent}, which admits programs up to GPU capacity and evicts programs by context length, regardless of each program's tool-call behavior.
  \item \textbf{Continuum}~\cite{li2025continuum}, ported to SGLang, which pins a program's KV cache in HBM during a tool call predicted to be short and unpins it when its TTL expires.
\end{packeditemize}

\MyPara{Methodology.} A closed-loop client keeps a fixed number of programs in flight, from 20 to 120. Each client replays one program at a time: it sends a request, waits for the response, sleeps for the recorded tool-call or human-wait duration, and sends the next request; when a program finishes, the client starts the next one. On AgentX, the subagents of a program run alongside it as additional programs. Each run lasts two hours. We report output throughput (output tokens per second over the run) and the p90 and median time to first token (TTFT) measured at the client, which includes queueing in the scheduler, KV cache reloads, and prefill.

\subsection{End-to-End Performance}
\label{sec:eval-main}

\Cref{fig:eval-main} shows throughput and p90 TTFT for every configuration and workload. \sys{} delivers the highest throughput and the lowest p90 TTFT at every point. At the highest concurrency, \sys{} improves throughput over the best baseline by 26--97\% (50\% on average) on TraceLab and 51--141\% (91\%) on AgentX, and lowers p90 TTFT by 56--89\% (72\%) and 66--76\% (71\%), respectively, relative to the baseline with the lowest p90; averaged over all nine configurations and workloads, the gains are 64\% and 72\%.

\MyPara{Scaling with load.} At 20 programs, the gain depends on how much KV cache the configuration holds. With Qwen2.5-7B, whose KV pool is the largest, every system serves at least 89\% of prompt tokens from cache, and \sys{} is at most 1\% ahead of the best baseline on TraceLab and 8\% ahead on AgentX. With gpt-oss-120b, whose pool is the smallest, the KV cache of 20 TraceLab programs already exceeds both tiers: Continuum and SGLang serve only 65--68\% of Codex prompt tokens from cache, and \sys{} leads by 8--21\% across the three workloads. Qwen3-235B lies in between ($+10$\% to $+14$\%). As concurrency grows, the aggregate KV cache exceeds both tiers in every configuration, and placement decides whose KV cache survives. The baselines flatten or decline, while \sys{} keeps scaling.

\MyPara{Across workloads.} The gains are largest on AgentX, followed by TraceLab Codex and TraceLab CC, because the workloads differ in how long programs stay idle (\Cref{fig:traces}(a)). Their gaps reach a p99 of 519--600\,s, and their contexts are long, so offloaded KV caches are large, must stay in DRAM for minutes, and cost much to recompute when dropped. AgentX has the fewest short gaps (44\% under 2\,s, versus 65--72\% in TraceLab), and its main agents also wait minutes for their subagents (median 3\,min). At 120 programs, ThunderAgent's hit rate thus falls to 63--79\% on TraceLab Codex and 34--68\% on AgentX, versus 92--95\% for \sys{} (\cref{sec:eval-why}). These heavy-tailed gaps also defeat Continuum's duration prediction (\cref{sec:mot-variance}): it pins only 6--36\% of tool calls on TraceLab, leaving the KV cache of the other programs to LRU. AgentX's traces do not name the tools, so this workload also emulates serving without tool information. \sys{} needs to know only which program a tool call belongs to, whereas Continuum must also know which tool is called: without tool names, its per-tool averages mix short and long calls and exceed its TTL, so it pins only 1--2\% of calls on AgentX. In TraceLab, Codex programs also spend longer in tool calls than CC programs (median gap 3.9\,s versus 0.09\,s), so more programs are idle at any time.

\MyPara{Latency.} \sys{}'s median TTFT stays below 2.5\,s at every point and its p90 below 50\,s; its median is at most 0.1\,s above ThunderAgent's because \sys{} reloads more programs from DRAM. ThunderAgent's tail degrades at high load: an idle program keeps its HBM until capacity overflows while ready programs wait in the scheduler queue (\cref{sec:mot-variance}), and the programs it evicts lose their KV cache, so at 120 programs its p90 reaches 730\,s on TraceLab and 2,307\,s on AgentX. SGLang and Continuum admit every request to the engine, where 89 to 125 requests wait on average at 120 programs. SGLang serves the longest cached prefix first, so requests whose KV cache was dropped wait at the back: with Qwen3-235B on TraceLab CC, they drive its p90 to 1,488\,s, while requests with cached prefixes see a p90 of 15\,s. Continuum serves requests in program-arrival order, which keeps its p90 below 190\,s, but even cached requests wait several seconds, and its median reaches 26\,s.

\subsection{Where the Gains Come From}
\label{sec:eval-why}

\begin{figure}[t]
  \centering
  \includegraphics[width=\columnwidth]{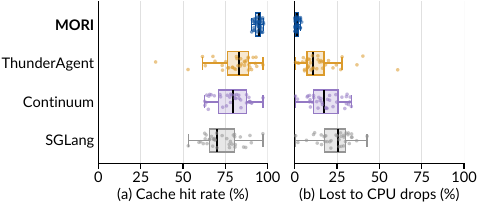}
  \caption{(a)~Fraction of prompt tokens served from the GPU or CPU cache, over all runs in \Cref{fig:eval-main}. (b)~Fraction of reusable prefix recomputed because its CPU copy had been dropped, over the same runs. Boxes span the 25th to 75th percentiles, black bars mark medians, whiskers extend to the furthest run within 1.5 times the interquartile range, and dots show individual runs.}
  \label{fig:eval-cache}
\end{figure}

At high load, most runs of every system keep the GPU above 90\% utilized, so throughput differences come mainly from how much GPU time goes to recomputing evicted context. \Cref{fig:eval-cache}(a) shows the fraction of prompt tokens served from the GPU or CPU cache. \sys{} serves 91--98\% (median 95\%), whereas the baselines' medians are 70--83\%, with lows of 34--63\%.

Much of the baselines' extra recomputation comes from the CPU tier (\Cref{fig:eval-cache}(b)). The median share of reusable prefix recomputed because its CPU copy had been dropped is 1.3\% for \sys{} but 11--26\% for the baselines. On AgentX, the baselines' medians rise to 17--32\% and their maxima to 32--61\%, while \sys{} stays below 4.5\%. The baselines offload every idle program and leave it to HiCache's LRU policy to drop CPU blocks when DRAM fills, so offloaded programs evict one another's KV cache, often shortly before they resume. \sys{} admits programs to DRAM only up to its capacity and keeps the rest in the Waiting queue, whose blocks both tiers evict first (\cref{sec:des-typed}), so it rarely drops a CPU copy that is needed again.

\subsection{GPU-to-CPU Capacity Ratio}
\label{sec:eval-ratio}
\begin{figure}[t]
  \centering
  \includegraphics[width=\columnwidth]{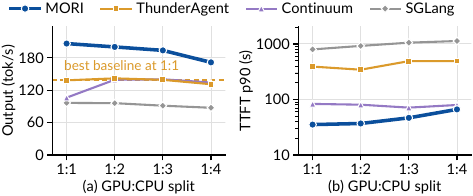}
  \caption{A fixed total KV capacity (2.59M tokens) split between GPU and CPU at GPU:CPU ratios from 1:1 to 1:4, for Qwen3-235B on TraceLab Codex with 80 programs (two B300 GPUs, TP=2). The 1:1 points are those of \Cref{fig:eval-main}.}
  \label{fig:eval-split-b}
\end{figure}

We also explore how \sys{} and the baselines adapt to different GPU-to-CPU ratios. In our default setup, the CPU holds as much KV cache as the GPU. To vary the ratio, we keep the total KV capacity unchanged at 2.59M tokens, as in \cref{sec:eval-main}, and divide it from GPU:CPU = 1:1 (1.29M tokens on the GPU and 1.30M on the CPU) to 1:4 (0.52M and 2.08M). We fix the total to ask how little HBM a system needs for the same performance: an idle program does no work whether its KV cache is in HBM or DRAM, so a good policy keeps only busy programs in HBM and needs less of it. We run Qwen3-235B on TraceLab Codex with 80 programs (\Cref{fig:eval-split-b}). Even at 1:4, where its GPU KV pool is only 40\% of the 1:1 size, \sys{} delivers 24\% more throughput than the best baseline does at 1:1 (dashed line in \Cref{fig:eval-split-b}(a)). The baselines thus use HBM inefficiently even at 1:1, leaving idle programs' KV cache in it (\cref{sec:mot-variance}), whereas \sys{} ranks programs by idleness and keeps only the most busy on the GPU (\cref{sec:des-policy}), so a smaller HBM still holds the programs that make the most use of it. \sys{} also remains the fastest system at every ratio, 28--49\% ahead of the best baseline, which is Continuum from 1:3 on as DRAM grows, and keeps its p90 TTFT at 35--66\,s, versus 71--83\,s for Continuum and 344--491\,s for ThunderAgent.

\section{Future Work}
\label{sec:future-work}

\MyPara{More offloading tiers.}\label{sec:disc-offloading}
\sys{} manages two tiers, GPU HBM and CPU DRAM, but its ranking extends to deeper hierarchies.
Serving nodes attach several terabytes of NVMe SSD, which is slower than DRAM but much larger and already serves as a KV cache tier~\cite{gao2024cachedattention}. The same ranking extends to it without any threshold: programs that do not fit in DRAM, typically the most idle ones in long tool calls, would go to SSD instead of the Waiting queue, just as programs that do not fit in HBM go to DRAM.
Offloading to SSD must also handle data corruption, as flash drives suffer uncorrectable errors far more often than ECC-protected DRAM~\cite{schroeder2016flash, schroeder2009dram}.
Remote DRAM on peer nodes, reachable over RDMA or NVMe-oF at 200\,Gb/s, is another tier between DRAM and SSD, as disaggregated KV cache systems show~\cite{qin2024mooncake, liu2025lmcache}.

\MyPara{Scheduling across replicas.}\label{sec:disc-lb}
With multiple data-parallel replicas, a router decides which replica serves each program. Routing must stay sticky, sending every request of a program to the replica that holds its KV cache. Within this constraint, idleness can drive load balancing: the router can give every replica a similar number of busy programs and thus a similar decode batch size.

\MyPara{Heterogeneous hardware.}\label{sec:disc-hetero}
Clusters increasingly mix accelerators and host DRAM capacities~\cite{mei2025helix, griggs2024melange}.
Because the idleness ranking is relative, replicas with different HBM sizes apply the same ranking; only where it is cut moves with local capacity.
The ranking could also route the most busy programs to the fastest replicas.

\section{Related Work}

\MyPara{LLM inference.}
Disaggregated prefill~\cite{zhong2024distserve, patel2024splitwise}, Preble~\cite{srivatsa2024preble}, and Llumnix~\cite{sun2024llumnix} schedule LLM inference at the cluster level. MemServe~\cite{hu2024memserve} unifies context caching with disaggregated inference through an elastic memory pool. vLLM lists hint-guided KV cache eviction and soft-pinning for agentic sessions as future work~\cite{vllm2026agentx}; \sys{} makes this program-level decision from request timing alone.

\MyPara{KV cache offloading.} FlexGen~\cite{sheng2023flexgen} places weights, activations, and KV cache across GPU, CPU, and disk for throughput-oriented batch inference, and Pie~\cite{xu2024piepoolingcpumemory} pools CPU memory with performance-transparent swapping. Subsequent systems target latency-sensitive online serving, where the reload cost of offloaded KV cache directly affects time-to-first-token~\cite{gao2024cachedattention, jiang2025neo, liu2025lmcache, liu2024cachegen, yao2024cacheblend, li2025economicalcontextaugmentedllmgeneration, qin2024mooncake}, or offload within a single long-context request~\cite{lee2024infinigen, sun2024shadowkv}. InferCept~\cite{abhyankar2024inferceptefficientinterceptsupport} decides per tool call whether to preserve, swap, or discard a request's KV cache. These systems build the hierarchy; \sys{} decides which agentic programs occupy each tier.

\MyPara{Agent paradigms.}
Early agents followed fixed workflows~\cite{shen2023hugginggpt, hong2023metagpt, qian2023chatdev}; as models grew more capable, ReAct-style agents, which let the LLM choose each next step, became the norm~\cite{yao2023react, shinn2023reflexion}. Multi-agent systems, despite known pitfalls~\cite{whymultiagentfail, guo2024multiagentsurvey}, are increasingly popular because they run subtasks in parallel and solve more complex tasks~\cite{li2026combeescalingpromptlearning, cusorparallelblog, ccparallelblog}.

\MyPara{Workflow-aware agent serving.} Parrot~\cite{lin2024parrot}, Teola~\cite{tan2025teola}, Ayo~\cite{tan2025ayo}, KVFlow~\cite{pan2025kvflow}, Helium~\cite{wadlom2026helium}, SAGA~\cite{guo2026saga}, and Pie~\cite{gim2025pie} assume that the workflow structure is fixed prior to agent execution and exploit it to schedule LLM calls and manage their KV cache, and PASTE~\cite{sui2026paste} speculatively executes predicted tool calls. Dynamic agents such as Claude Code (CC)~\cite{anthropic2025claudecode}, Codex~\cite{openai2025codex}, and Cursor~\cite{anysphere2024cursor}, however, decide each next step at run time, so their workflow is not known in advance. Autellix~\cite{luo2025autellix} orders calls by per-program attained service, without managing KV retention.

\section{Conclusion}

This paper studied program eviction and promotion for agentic LLM workloads, whose tool-call durations vary by orders of magnitude and are unknown in advance. Offloading to CPU DRAM lets a program wait out a long tool call instead of recomputing its context, but existing program-aware schedulers manage only HBM and leave DRAM to LRU, so offloaded programs thrash. Our trace analysis shows that programs stay in busy or idle phases for extended periods, and because the busy/idle split need not match the GPU-to-CPU capacity ratio, what matters is how idle a program is relative to others. \sys{} ranks programs by idleness, their recent fraction of time in tool calls, keeps the least idle in HBM and the next in DRAM, and drops the rest, enforcing admission control at both tiers; it also promotes ready programs proactively. Under high load on the three coding-agent workloads, \sys{} achieves on average 64\% higher throughput and 72\% lower p90 TTFT than the best baseline, and with the same total KV capacity, it delivers 24\% more throughput than the best baseline while moving 60\% of the GPU KV to CPU.

\section*{Acknowledgement}
This work is supported by gifts from Accenture, Algorithmic SuperIntelligence Labs, Amazon, AMD, Anyscale, Broadcom, cmpnd, Google, IBM, Intel, Intesa Sanpaolo, Lambda, Lightspeed, Mirendil, NVIDIA, Soma Labs, and VESSL.

\bibliographystyle{mlsys2025}
\bibliography{sample-base}

\appendix

\end{document}